\documentclass[journal]{IEEEtran}
\usepackage{amsmath}
\usepackage{amssymb}
\usepackage{algorithm}
\usepackage{algorithmicx}
\usepackage{algpseudocode}
\usepackage[T1]{fontenc}
\usepackage{color}
\usepackage{times}
\usepackage{cite}

\usepackage{graphicx}
\usepackage{float} 
\usepackage{subfigure}
\usepackage{setspace}
\usepackage{stfloats}

\begin{document}
	\newcounter{TempEqCnt}
\title{\huge How Much Time is Required for Phase Shift Delivery in RIS-Aided Wireless Systems?}
\author{Hao Xie and Dong Li,~\IEEEmembership{Senior Member,~IEEE} \vspace{-30pt}
	\thanks{H. Xie and D. Li are with the School of Computer Science and Engineering, Macau University of Science and Technology, Avenida Wai Long, Taipa, Macau 999078, China (e-mails: 3220005631@student.must.edu.mo, dli@must.edu.mo).}
} 

\maketitle

\begin{abstract}
Reconfigurable intelligent surface (RIS) has become a focal point of extensive research due to its remarkable ``squared gain''. However, achieving a substantial beamforming gain typically requires a significant number of elements, which leads to a non-negligible overhead that forwards the coherent phase shift to the RIS. Different from previous works, which primarily focus on the information transmission phase, we consider the phase delivery overhead during the phase-shift delivery phase to explore the trade-off between performance and overhead. To reduce the phase delivery overhead via the control link, we introduce a hybrid phase shift mechanism, encompassing both the coherent and fixed phase shifts. Specifically, a beamforming problem is formulated for maximizing the throughput. In light of the intractability of the problem, we develop an alternating optimization-based iterative algorithm by combining quadratic transformation and successive convex approximation. To gain more insights, we derive the closed-form expression of the number of elements adopting the coherent phase shift in the large signal-to-noise ratio region. This expression serves as a valuable guide for the practical implementation of the RIS technology. Our simulation results conclusively demonstrate the effectiveness of the proposed algorithm in achieving a favorable trade-off between throughput and overhead.  Furthermore, the introduction of the hybrid phase shift approach significantly reduces phase delivery overhead while concurrently enhancing the system throughput.
\end{abstract}



\vspace{-10pt}
\section{Introduction}
Recently, reconfigurable intelligent surface (RIS) has garnered substantial attention from both academia and industry due to its remarkable attributes including high array gain and low power consumption \cite{a1}. The RIS can passively reflect the received signal and achieve ``squared gain'' without extra transmit power and channel bandwidth resources by simply adjusting the phase shift of each reflecting element. So far, there have been numerous efforts to apply the RIS to various network scenarios, such as mmWave communication \cite{b1}, wireless-powered communication\cite{b2},  and to improve the performance and efficiency of communication systems by optimizing different performance metrics, such as the age of information \cite{c1}, energy efficiency\cite{c2}, deployment location\cite{c3}. 

However, it is worth noting that current research primarily focuses on information transmission, overlooking a crucial aspect: the phase-shift delivery, which typically refers to the base station (BS) delivering the optimized phase shift to the RIS so that the passive beamforming can be achieved. In fact, the duration of the phase-shift delivery process primarily depends on the phase delivery overhead and has a direct influence on the duration of the subsequent information transmission phase. Up to now, the phase-shift delivery process has received little attention, and there are only a few works (see, e.g., \cite{d1,d2,d3})  related to this work regarding the delivery overhead. However, the trade-off between the delivery time and the transmission time and how to configure the number of reflecting elements are not covered in existing works \cite{d1,d2,d3}. Specifically, although a substantial number of elements can improve throughput performance, it also reduces the transmission time due to the prolonged phase delivery, and thus deteriorating the transmission performance. Thus, a natural question arises: \textit{How long is required for phase shift delivery?} This question is generally neglected in existing works where the phase delivery is not involved, and this is the first attempt to solve this problem to the best of the authors' knowledge.


%
%

Inspired by the aforementioned observations, we investigate the information transmission with phase delivery in RIS-aided wireless communications. We carefully examine the trade-off between delivery time and transmission time by investigating the hybrid phase shift mechanism. This hybrid phase shift combines both coherent and fixed phase shifts, which empowers the system to achieve a delicate equilibrium between the system performance and the delivery overhead. Specifically, we formulate an optimization problem aiming at maximizing the throughput by taking the hybrid phase shift and time slot for phase delivery into account. Recognizing the complexity inherent in the formulated problem, we propose an alternating optimization-based algorithm that adopts quadratic transformation and the successive convex approximation (SCA) method to obtain the sub-optimal solution. To gain deeper insights, we explore the number of elements adopting coherent phase shifts and derive a closed-form expression in the large signal-to-noise ratio (SNR) region. This information can serve as a practical guide for the deployment of the RIS. Our simulation results substantiate that the proposed optimization framework can effectively reduce the phase delivery overhead and enhance the throughput.

\vspace{-10pt}
\section{System Model and Problem Formulation}
\begin{figure}[!t]\vspace{-25pt}
	\centering
	\includegraphics[width=3in]{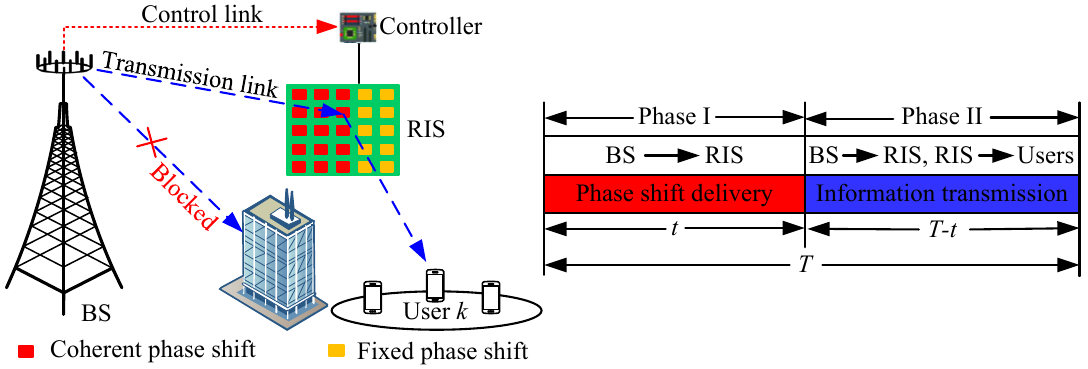}
	\caption{The RIS-aided wireless communications.}
	\label{fig0}\vspace{-25pt}
\end{figure}

As illustrated in Fig. \ref{fig0}, we consider a RIS-aided wireless communication network consisting of a BS equipped with $M$ antennas, a RIS equipped with $N$ reflecting elements, and $K$ single-antenna users. The RIS is equipped with a controller with signal processing and transmission/reception capabilities, harmonizing the interactions between the BS and the RIS for precise phase shift control. However, delivering the optimized phase shift to the RIS before the data transmission phase will have an impact on the communication phase, especially for a large number of reflecting elements. To this end, we consider a hybrid phase shift mechanism that includes both the coherent phase shift\cite{e1}, which needs to be optimized and delivered to the controller, and the fixed phase shift, which is generated randomly at the RIS without any phase delivery overhead. The RIS comprises two co-located sub-surfaces with $N^{\rm coh}$ coherent phase shifts and $N^{\rm fix}$ fixed phase shifts. Note that the hybrid phase shift can be reduced to the traditional coherent phase shift (see, e.g., \cite{e1a}) when $N^{\rm fix}=0$. Define $\theta_{n}^{\rm coh}$ and $\theta_{n}^{\rm fix}$ as coherent phase shift and fixed phase shift, respectively. Thus, the phase-shift matrix can be denoted as $\boldsymbol{\rm \Theta}={\textrm{diag}}(e^{j\theta_{1}^{\rm coh}},\cdots,e^{j\theta_{N^{\rm coh}}^{\rm coh}},e^{j\theta_{1}^{\rm fix}},\cdots, e^{j\theta_{N^{\rm fix}}^{\rm fix}})$ and $N=N^{\rm coh}+N^{\rm fix}$.
Specifically, we denote $\boldsymbol{\rm \Theta}^{\rm coh}{=}{\textrm{diag}}(e^{j\theta_{1}^{\rm coh}},\cdots,e^{j\theta_{n}^{\rm coh}},\cdots,e^{j\theta_{N^{\rm coh}}^{\rm coh}})$ as the phase-shift matrix for the coherent phase shift and denote $\boldsymbol{\rm \Theta}^{\rm fix}={\textrm{diag}}(e^{j\theta_{1}^{\rm fix}},\cdots,e^{j\theta_{n}^{\rm fix}},\cdots,e^{j\theta_{N^{\rm fix}}^{\rm fix}})$ as the phase-shift matrix for the fixed phase shift. 
For the phase-shift delivery phase, we define $b$ as the number of quantization bit for each coherent phase shift and define $t$ as the duration of the phase delivery via the control link. Then, the delivery bits should satisfy the following condition
\begin{equation}\label{eq1}
\begin{array}{l}
tR_{\rm F}\geq bN^{\rm coh},
\end{array}
\end{equation}
where $R_{\rm F}$ denotes the delivery rate needed in the control link, which can be obtained by the method in \cite{d2}. In the information transmission phase, the BS transmits information to $K$ users with the assistance of the RIS. Based on the hybrid phase shift mechanism, the received signal at the $k$-th user is\cite{b1}
\begin{equation}\label{eq2}
\begin{array}{l}
\!\!\!\!\!\!y_k
=\sum\limits_{k=1}^K\boldsymbol{\rm h}_{{\rm r}, k}^H\boldsymbol{\rm \Theta}\boldsymbol{\rm H}\boldsymbol{\rm w}_ks_k+n_k=\sum\limits_{k=1}^K((\boldsymbol{\rm h}_{{\rm r}, k}^{\rm coh})^H\boldsymbol{\rm \Theta}^{\rm coh}\boldsymbol{\rm H}^{\rm coh}\\
+(\boldsymbol{\rm h}_{{\rm r}, k}^{\rm fix})^H\boldsymbol{\rm \Theta}^{\rm fix}\boldsymbol{\rm H}^{\rm fix})\boldsymbol{\rm w}_ks_k+n_k,
\end{array}
\end{equation}
where $\boldsymbol{\rm w}_k$ and $s_k$ denote the beamforming vector and the transmission signal with $\mathbb{E}\{|s_k|^2\}=1$ of the $k$-th user, respectively. $(\cdot)^H$ denotes the conjugate transpose operator. $\boldsymbol{\rm h}_{{\rm r},k}\in\mathbb{C}^{N\times 1}=\begin{bmatrix}\boldsymbol{\rm h}_{{\rm r},k}^{\rm coh};\boldsymbol{\rm h}_{{\rm r},k}^{\rm fix}
\end{bmatrix}$ and  $\boldsymbol{\rm H}\in\mathbb{C}^{N\times M}=\begin{bmatrix}\boldsymbol{\rm H}^{\rm coh};\boldsymbol{\rm H}^{\rm fix}
\end{bmatrix}$ denote the channels from the RIS to the $k$-th user and the BS to the RIS, respectively. $n_k\sim\mathcal{CN}(0,\delta_k^2)$ denotes the additive white Gaussian noise (AWGN) at the $k$-th user.
Thus, the signal-to-interference-plus-noise ratio (SINR) of  user $k$ is
	\begin{equation}\label{eq3}
	\begin{array}{l}
	\gamma_k=\frac{|((\boldsymbol{\rm h}_{{\rm r}, k}^{\rm coh})^H\boldsymbol{\rm \Theta}^{\rm coh}\boldsymbol{\rm H}^{\rm coh}+(\boldsymbol{\rm h}_{{\rm r}, k}^{\rm fix})^H\boldsymbol{\rm \Theta}^{\rm fix}\boldsymbol{\rm H}^{\rm fix})\boldsymbol{\rm w}_k|^2}{\sum\limits_{i\not=k}^K|((\boldsymbol{\rm h}_{{\rm r}, k}^{\rm coh})^H\boldsymbol{\rm \Theta}^{\rm coh}\boldsymbol{\rm H}^{\rm coh}+(\boldsymbol{\rm h}_{{\rm r}, k}^{\rm fix})^H\boldsymbol{\rm \Theta}^{\rm fix}\boldsymbol{\rm H}^{\rm fix})\boldsymbol{\rm w}_i|^2+\delta_k^2}.
	\end{array}
	\end{equation}
Then, the achievable throughput of the $k$-th user is formulated as
$
R_k=(T-t)\log_2(1+\gamma_k)
$,
where $T$ denotes the total duration of phase delivery and information transmission phase.

In this paper, our objective is to find a harmonious equilibrium between the throughput and delivery overhead in RIS-aided wireless networks. To be more specific, we formulate the problem of throughput maximization as follows
\begin{equation}\label{eq4}
\begin{array}{l}
\max\limits_{\mbox{\scriptsize$\begin{array}{c} 
		\boldsymbol{\rm \Theta}^{\rm coh},\boldsymbol{\rm w}_k,t
		\end{array}$}} 
\sum\limits_{k=1}^K(T-t)\log_2(1+\gamma_k) \\
s.t.~
{C_1}:pt+(T-t)\sum\limits_{k=1}^K\|\boldsymbol{\rm w}_k\|^2\leq E^{\max},\\
~~~~~{C_2}:(T-t)\log_2(1+\gamma_k)\geq R_k^{\min},\\
~~~~~{C_3}:|[\boldsymbol{\rm \Theta}^{\rm coh}]_{n,n}|=1,n=1,\cdots,N^{\rm coh},\\
~~~~~{C_4}:tR_{\rm F}\geq bN^{\rm coh},
\end{array}
\end{equation}
where $p$ denotes the delivery power,
$E^{\max}$ represents the maximum energy threshold of the BS, $R_k^{\min}$ signifies the minimum transmission throughput threshold for user $k$. $C_1$ corresponds to the maximum energy constraint enforced on the BS, $\sum_{k=1}^K\|\boldsymbol{\rm w}_k\|^2$ denotes the sum power of the BS\cite{c1}. $C_2$ represents the minimum throughput constraint imposed on the $k$-th user, guaranteeing a specified level of data transmission performance. $C_3$ represents phase shifts $\theta_n^{\rm coh}$ changing within the $[0, 2\pi)$. For ease of handling, we have expressed it in the equivalent form mentioned above\cite{c3}. $C_4$ denotes the constraint on delivery bits. It is noted that the formulated problem is completely different from the existing works \cite{d1,d2,d3} due to the introduction of the hybrid phase shift and phase delivery overhead, which results in different solution methodologies, and the algorithms designed in existing works cannot be applied in our work.

\vspace{-10pt}
\section{Beamforming Algorithm Design}
The main challenge in problem (\ref{eq4}) lies in the introduction of phase shift delivery, exacerbating the coupling among variables. Besides, the fractional form in the rate expression renders problem (\ref{eq4}) challenging to solve. Conventional fractional programming techniques, such as the Charnes-Cooper method and Dinkelbach’s method, perform effectively in single-ratio scenarios but encounter challenges when addressing multi-ratio cases. To this end, we employ quadratic transformation proposed in \cite{f1} to decouple the variables in problem (\ref{eq4}), as demonstrated in the following Lemma.


\textbf{\textit{Lemma 1:}} (Equivalent problem for throughput maximization): Through the introduction of auxiliary variables $\rho_k$ and $\eta_k$, we can equivalently reformulate the original problem presented in (\ref{eq4}) as follows
\begin{equation}\label{eq5}
\begin{array}{l}
\max\limits_{\mbox{\scriptsize$\begin{array}{c} 
		\boldsymbol{\rm \Phi}^{\rm coh},
		\boldsymbol{\rm w}_k,t,\rho_k,\eta_k
		\end{array}$}} 
R_{\rm sum}=(T-t)(\sum\limits_{k=1}^K\log_2(1+\rho_k)\\
~~~~~~~~~~~~~~~~~~~~~~~~~~~~-\sum\limits_{k=1}^K\rho_k+\sum\limits_{k=1}^Kf_k)\\
~~~~~~~~~~~~~~~~~s.t.~
{C_1},C_2,{C_3},C_4,
\end{array}
\end{equation}
where
\begin{equation}\label{eq6}
\begin{array}{l}
f_k=\frac{(1+\rho_k)\gamma_k}{1+\gamma_k}=2\eta_k\times\\
\sqrt{(1+\rho_k)|((\boldsymbol{\rm h}_{{\rm r}, k}^{\rm coh})^H\boldsymbol{\rm \Theta}^{\rm coh}\boldsymbol{\rm H}^{\rm coh}+(\boldsymbol{\rm h}_{{\rm r}, k}^{\rm fix})^H\boldsymbol{\rm \Theta}^{\rm fix}\boldsymbol{\rm H}^{\rm fix})\boldsymbol{\rm w}_k|^2}\\
-\eta_k^2(\sum\limits_{i=1}^K|((\boldsymbol{\rm h}_{{\rm r}, k}^{\rm coh})^H\boldsymbol{\rm \Theta}^{\rm coh}\boldsymbol{\rm H}^{\rm coh}+(\boldsymbol{\rm h}_{{\rm r}, k}^{\rm fix})^H\boldsymbol{\rm \Theta}^{\rm fix}\boldsymbol{\rm H}^{\rm fix})\boldsymbol{\rm w}_i|^2
+\delta_k^2).
\end{array}
\end{equation}
\begin{proof}
	please refer to \cite{f1}.
\end{proof}

Once the other variables are held constant, the optimal value for $\rho_k$ can be obtained by solving the equation $\frac{\partial R_{\rm sum}}{\partial \rho_k}=0$, i.e., $\rho_k=\gamma_k$.
After fixing other variables, the optimal $\eta_k$ can be obtained by solving $\frac{\partial R_{\rm sum}}{\partial \eta_k}=0$, i.e.,
\begin{equation}\label{eq7}
\begin{array}{l}
\eta_k=\frac{\sqrt{(1+\rho_k)|((\boldsymbol{\rm h}_{{\rm r}, k}^{\rm coh})^H\boldsymbol{\rm \Theta}^{\rm coh}\boldsymbol{\rm H}^{\rm coh}+(\boldsymbol{\rm h}_{{\rm r}, k}^{\rm fix})^H\boldsymbol{\rm \Theta}^{\rm fix}\boldsymbol{\rm H}^{\rm fix})\boldsymbol{\rm w}_k|^2}}{\sum\limits_{i=1}^K|((\boldsymbol{\rm h}_{{\rm r}, k}^{\rm coh})^H\boldsymbol{\rm \Theta}^{\rm coh}\boldsymbol{\rm H}^{\rm coh}+(\boldsymbol{\rm h}_{{\rm r}, k}^{\rm fix})^H\boldsymbol{\rm \Theta}^{\rm fix}\boldsymbol{\rm H}^{\rm fix})\boldsymbol{\rm w}_i|^2+\delta_k^2},
\end{array}
\end{equation}

Next, we will iteratively optimize $\boldsymbol{\rm w}_k$, $t$, and $	\boldsymbol{\rm \Theta}^{\rm coh}$. The specific optimization process is shown as follows.

\subsubsection{Delivery Time Optimization}
In this subsection, we focus on the delivery time optimization for given other variables. 
Then, we can transform problem (\ref{eq5}) into
\begin{equation}\label{eq8}
\begin{array}{l}
\min\limits_{\mbox{\scriptsize$\begin{array}{c} 
		t
		\end{array}$}} 
~~~~~~t\\
s.t.~
{C_1},{C_2},{C_4}.
\end{array}
\end{equation}
It is important to highlight that problem (\ref{eq8}) is a linear programming problem with respect to the variable $t$. Such linear programming problems can be efficiently solved using dedicated optimization tools, e.g., linprog.

\subsubsection{Beamforming Vector Optimization}
In this subsection, our focus lies in optimizing the beamforming vector while keeping the other variables fixed. Consequently, we can transform problem (\ref{eq5}) into the following form
\begin{equation}\label{eq9}
\begin{array}{l}
\max\limits_{\mbox{\scriptsize$\begin{array}{c} 
		\boldsymbol{\rm w}_k
		\end{array}$}} 
R_{\rm sum} \\
s.t.~
{C_1},{C_2}.
\end{array}
\end{equation}
To begin with, we introduce a new variable $\boldsymbol{\rm W}_k=\boldsymbol{\rm w}_k\boldsymbol{\rm w}_k^H$. This transformation allows us to express the following
\begin{equation}\label{eq10}
\begin{array}{l}
\bar C_1: pt+(T-t)\sum\limits_{k=1}^K{\rm Tr}(\boldsymbol{\rm W}_k)\leq E^{\max},\\
\bar C_2: {\rm Tr}(\boldsymbol{\rm H}_{k}\boldsymbol{\rm W}_k)\\
~~~~~-(2^{\frac{R_k^{\min}}{T-t}}-1)(\sum\limits_{i\not=k}^K{\rm Tr}(\boldsymbol{\rm H}_{k}\boldsymbol{\rm W}_i)+\delta_k^2)\geq 0,\\
\bar f_k=2\eta_k\sqrt{(1+\rho_k){\rm Tr}(\boldsymbol{\rm H}_{k}\boldsymbol{\rm W}_k)}\\
~~~~~~~~~~~~~~~~~~~~~~~-\eta_k^2(\sum\limits_{i=1}^K{\rm Tr}(\boldsymbol{\rm H}_{k}\boldsymbol{\rm W}_i)+\delta_k^2),
\end{array}
\end{equation}
where $\boldsymbol{\rm H}_{k}=\boldsymbol{\rm h}_{k}\boldsymbol{\rm h}_{k}^H$ and $\boldsymbol{\rm h}_{k}=(\boldsymbol{\rm h}_{{\rm r}, k}^{\rm coh})^H\boldsymbol{\rm \Theta}^{\rm coh}\boldsymbol{\rm H}^{\rm coh}+(\boldsymbol{\rm h}_{{\rm r}, k}^{\rm fix})^H\boldsymbol{\rm \Theta}^{\rm fix}\boldsymbol{\rm H}^{\rm fix}$. It is essential to emphasize that the introduction of the new variable helps transform the problem into a convex one, but it also brings a non-convex equality constraint in the form of $\boldsymbol{\rm W}_k=\boldsymbol{\rm w}_k\boldsymbol{\rm w}_k^H$. In order to address this, we equivalently transform the equality constraint into the following set of constraints\cite{c2}
\begin{equation}\label{eq13}
\begin{array}{l}
\!\!\!\!\!C_{5a}\!:\!\left[                 
\begin{array}{cc}   
\boldsymbol{\rm W}_k& \boldsymbol{\rm w}_k\\  
\boldsymbol{\rm w}_k^H & 1\\  
\end{array}
\right] \succeq \boldsymbol{0},C_{5b}\!:\! {\rm Tr}(\boldsymbol{\rm W}_k-\boldsymbol{\rm w}_k\boldsymbol{\rm w}_k^H)\leq 0,
\end{array}     
\end{equation}
Furthermore, it is worth noting that $C_{5b}$ remains a non-convex constraint. Leveraging the SCA method, we can derive its lower bound as
${\rm Tr}(\boldsymbol{\rm w}_k\boldsymbol{\rm w}_k^H)\geq -\|\boldsymbol{\rm w}_k^{(l)}\|^2+2{\rm Tr}((\boldsymbol{\rm w}_k^{(l)})^H\boldsymbol{\rm w}_k)$,
where $\boldsymbol{\rm w}_k^{(l)}$ is the $l$-th iteration of $\boldsymbol{\rm w}_k$. Thus, by substituting the lower bound into $C_{5b}$, $C_{5b}$ can be rewritten as
\begin{equation}\label{eq15}
\begin{array}{l}
\bar C_{5b}:{\rm Tr}(\boldsymbol{\rm W}_k)\leq -\|\boldsymbol{\rm w}_k^{(l)}\|^2+2{\rm Tr}((\boldsymbol{\rm w}_k^{(l)})^H\boldsymbol{\rm w}_k),
\end{array}
\end{equation}
Then, problem (\ref{eq9}) can be transformed into the following convex problem
\begin{equation}\label{eq16}
\begin{array}{l}
\!\!\!\!\!\!\!\max\limits_{\mbox{\scriptsize$\begin{array}{c} 
		\boldsymbol{\rm w}_k,\boldsymbol{\rm W}_k
		\end{array}$}} 
(T-t)(\sum\limits_{k=1}^K\log_2(1+\rho_k)-\sum\limits_{k=1}^K\rho_k+\sum\limits_{k=1}^K\bar f_k) \\
~~~s.t.~
{\bar C_1},{\bar C_2}, C_{5a}, \bar C_{5b},C_6:\boldsymbol{\rm W}_k\succeq \boldsymbol{0}.
\end{array}
\end{equation}
\subsubsection{Phase Shift Optimization}
In this subsection, we optimize the phase shift for given other variables. Define $\boldsymbol{\rm v}=[e^{j\theta_1^{\rm coh}},\cdots,e^{j\theta_{N^{\rm coh}}^{\rm coh}}]^T$, then problem (\ref{eq5}) can be reduced to
\begin{equation}\label{eq17}
\begin{array}{l}
\max\limits_{\mbox{\scriptsize$\begin{array}{c} 
		\boldsymbol{\rm \Phi}^{\rm coh}
		\end{array}$}} 
R_{\rm sum} \\
s.t.~
{C_2},C_3.
\end{array}     
\end{equation}
To facilitate algorithm design, we have the following transformation: $|((\boldsymbol{\rm h}_{{\rm r}, k}^{\rm coh})^H\boldsymbol{\rm \Theta}^{\rm coh}\boldsymbol{\rm H}^{\rm coh}+(\boldsymbol{\rm h}_{{\rm r}, k}^{\rm fix})^H\boldsymbol{\rm \Theta}^{\rm fix}\boldsymbol{\rm H}^{\rm fix})\boldsymbol{\rm w}_k|^2
={\rm Tr}(\boldsymbol{\rm W}_k\boldsymbol{\rm H}_{1,k}^H\boldsymbol{\rm O}\boldsymbol{\rm H}_{1,k})
$, where $\boldsymbol{\rm O}=\boldsymbol{\rm o}\boldsymbol{\rm o}^H$ and $\boldsymbol{\rm o}^H=[\boldsymbol{\rm v}^H,1]$. $\boldsymbol{\rm H}_{1,k}=[{\rm diag}( (\boldsymbol{\rm h}_{{\rm r}, k}^{\rm coh})^H )\boldsymbol{\rm H}^{\rm coh};(\boldsymbol{\rm h}_{{\rm r}, k}^{\rm fix})^H\boldsymbol{\rm \Theta}^{\rm fix}\boldsymbol{\rm H}^{\rm fix}]$. Similar to (\ref{eq13}), the equality constraint $\boldsymbol{\rm O}{=}\boldsymbol{\rm o}\boldsymbol{\rm o}^H$ can be transformed into
\begin{equation}\label{eq18}
\begin{array}{l}
C_{7a}:\left[                 
\begin{array}{cc}   
\boldsymbol{\rm O}& \boldsymbol{\rm o}\\  
\boldsymbol{\rm o}^H & 1\\  
\end{array}
\right] \succeq \boldsymbol{0},\\
C_{7b}: {\rm Tr}(\boldsymbol{\rm O})\leq -\|\boldsymbol{\rm o}^{(l)}\|^2+2{\rm Tr}((\boldsymbol{\rm o}^{(l)})^H\boldsymbol{\rm o}),\\
\end{array}     
\end{equation}
where $\boldsymbol{\rm o}^{(l)}$ is the $l$-th iteration of $\boldsymbol{\rm o}$. Then, problem (\ref{eq17}) can be transformed into the following convex problem
\begin{equation}\label{eq19}
\begin{array}{l}
\!\!\!\!\max\limits_{\mbox{\scriptsize$\begin{array}{c} 
		\boldsymbol{\rm O}, \boldsymbol{\rm o}
		\end{array}$}} 
~~(T-t)(\sum\limits_{k=1}^K\log_2(1+\rho_k)-\sum\limits_{k=1}^K\rho_k+\sum\limits_{k=1}^K \hat f_k) \\
s.t.~
\hat C_3:[\boldsymbol{\rm O}]_{n,n}\leq 1,\boldsymbol{\rm O}\succeq\boldsymbol{0},[\boldsymbol{\rm O}]_{N+1,N+1}= 1,\\
~~~~~\hat C_2, C_{7a}, C_{7b},
\end{array}
\end{equation}
where $\hat f_k=2\eta_k\sqrt{(1+\rho_k){\rm Tr}(\boldsymbol{\rm O}\boldsymbol{\rm H}_{1,k}\boldsymbol{\rm W}_k\boldsymbol{\rm H}_{1,k}^H)}-\eta_k^2(\sum_{i=1}^K\\{\rm Tr}(\boldsymbol{\rm O}\boldsymbol{\rm H}_{1,k}\boldsymbol{\rm W}_i\boldsymbol{\rm H}_{1,k}^H){+}\delta_k^2)$ and $\hat C_2: {\rm Tr}(\boldsymbol{\rm O}\boldsymbol{\rm H}_{1,k}\boldsymbol{\rm W}_k\boldsymbol{\rm H}_{1,k}^H)-(2^{\frac{R_k^{\min}}{T-t}}-1)(\sum_{i\not=k}^K{\rm Tr}(\boldsymbol{\rm O}\boldsymbol{\rm H}_{1,k}\boldsymbol{\rm W}_i\boldsymbol{\rm H}_{1,k}^H)+\delta_k^2)\geq 0$.

\section{How many reflecting elements with coherent phase shift should be used?}
In the previous section, we have formulated a general optimization problem to analyze the trade-off between phase delivery time and transmission time. In addition to general algorithm design, it is also imperative to provide specific guidance for the practical deployment of the RIS and delivery time in the system, which promotes the development of this section. With the aim of gaining deeper insights, we explore the number configuration of the coherent phase shift and analyze how much delivery time is needed to enhance the throughput. Note that the number configuration and the topic of this work are different from previous works (see, e.g., \cite{e1,e1a,f3}) in the sense that the impact of both the phase delivery and the number of coherent phase shift on the data transmission is investigated in this work for the first time. To facilitate our analysis, we consider a simplified communication scenario, where the BS equipped with a single antenna serves one user and performs information transmission at a fixed power level $p_{\rm t}$, all reflecting elements adopt the coherent phase shift. Consequently, problem (\ref{eq4}) is simplified to the following form
\begin{equation}\label{eq20}
\begin{array}{l}
\max\limits_{\mbox{\scriptsize$\begin{array}{c} 
		\boldsymbol{\rm \Theta}^{\rm coh},t
		\end{array}$}} 
(T-t)\log_2(1+\frac{p_{\rm t}|(\boldsymbol{\rm h}_{{\rm r}}^{\rm coh})^H\boldsymbol{\rm \Theta}^{\rm coh}\boldsymbol{\rm h}^{\rm coh}|^2}{\delta^2}) \\
s.t.~
{C_1}:pt+(T-t)p_{\rm t}\leq E^{\max},\\
~~~~~{C_2}:(T-t)\log_2(1+\frac{p_{\rm t}|(\boldsymbol{\rm h}_{{\rm r}}^{\rm coh})^H\boldsymbol{\rm \Theta}^{\rm coh}\boldsymbol{\rm h}^{\rm coh}|^2}{\delta^2})\geq R^{\min},\\
~~~~~{C_3}:|[\boldsymbol{\rm \Theta}^{\rm coh}]_{n,n}|=1,\forall n=1,\cdots,N^{\rm coh},\\
~~~~~{C_4}:tR_{\rm F}\geq bN^{\rm coh},
\end{array}
\end{equation}
It is worth noting that problem (\ref{eq20}) is a simplified version of problem (\ref{eq4}), and the optimizations of these two problems are independent of each other.
We observe that the delivery time has an effect on the throughput, i.e., a shorter delivery time yields a higher throughput. Without loss of generality, we assume $p_{\rm t}>p$. Then, we can find from the optimization problem (\ref{eq20}) that the objective function is a monotonically decreasing function with respect to $t$. $C_1$ and $C_4$ are lower-bound constraints on $t$, and $C_2$ is an upper-bound constraint on $t$. Thus, we have 
\begin{equation}\label{eq22}
\begin{array}{l}
t^*=\max\{\frac{bN^{\rm coh}}{R_{\rm F}},\frac{E^{\max}-Tp_{\rm t}}{p-p_{\rm t}}\}.
\end{array}
\end{equation}
It is worth noting that $t^*$ must satisfy the constraint $(T-t^*)\log_2(1+\gamma)\geq R^{\min}$, otherwise the problem (\ref{eq20}) is infeasible. Next, we first drop $C_2$ to make problem (\ref{eq20}) more treatable and then analyze its feasible conditions at the end. In the following, we will consider two cases to analyze the number configuration for the coherent phase shift.
\subsubsection{Case 1 with $t^*=\frac{bN^{\rm coh}}{R_{\rm F}}$} In order to promote the algorithm design, we first deal with the SNR expression, then we have the following inequation
\begin{equation}\label{eq23}
\begin{array}{l}
\dfrac{p_{\rm t}|(\boldsymbol{\rm h}_{{\rm r}}^{\rm coh})^H\boldsymbol{\rm \Theta}^{\rm coh}\boldsymbol{\rm h}^{\rm coh}|^2}{\delta^2}\overset{(a)}{=}\dfrac{p_{\rm t}\left|\sum\limits_{n=1}^{N^{\rm coh}} |h_{{\rm r},n}^{\rm coh}| |h_n^{\rm coh}|\right|^2}{\delta^2}\\
~~~~~~~~~~~~~~~~~~~~~~~~~~~\overset{(b)}{\geq} \dfrac{p_{\rm t} |h_{{\rm r},n}^{\rm coh}|^2 |h_n^{\rm coh}|^2 (N^{\rm coh})^2}{\delta^2},\\
\end{array}
\end{equation}
where $(a)$ utilizes the optimal design of $\boldsymbol{\rm \Theta}^{\rm coh}$, $(b)$ utilizes $|h_{\rm r}^{\rm coh}|=\min\{|h_{{\rm r},n}^{\rm coh}|\}$ and $|h^{\rm coh}|=\min\{|h_{n}^{\rm coh}|\}$. Then, when $t=\frac{bN^{\rm coh}}{R_{\rm F}}$ holds, we have the following proposition:

\textit{Proposition 1:} In the large SNR region, i.e., $\bar p=\frac{p_{\rm t}}{\delta^2}\to \infty$, when $N^{\rm coh}>0$, define $C=|h_{\rm r}^{\rm coh}|^2|h^{\rm coh}|^2$, we can approximate the closed-form solution for $N^{\rm coh}$ as follows
\begin{equation}\label{eq24}
\begin{array}{l}
N^{\rm coh,*}=\frac{TR_{\rm F}}{bW_0\left(\frac{eTR_{\rm F}\sqrt{\bar p C}}{b}\right)},
\end{array}
\end{equation}
where $W_0(\cdot)$ denotes the principal branch of the Lambert’s $W$ function. The optimization of the number of coherent phase shifts is derived through optimizing the phase shifts and is not directly performed when designing problem (\ref{eq20}).
\begin{proof}
	As $\bar p\to \infty$, we can obtain that $\log_2(1+a\bar p)=\log_2(a\bar p)+o(1)$. Thus, the objective function can be approximated as
	\begin{equation}\label{eq25}
	\begin{array}{l}
	f(N^{\rm coh})\triangleq(T-\frac{bN^{\rm coh}}{R_{\rm F}})\log_2(\bar pC(N^{\rm coh})^2)+o(1).
	\end{array}
	\end{equation}
	Considering the leading-order term and taking the derivative of (\ref{eq25}) with respect to $N^{\rm coh}$ equaling to 0, we can obtain
	\begin{equation}\label{eq26}
	\begin{array}{l}
	\frac{TR_{\rm F}}{N^{\rm coh}}=b+b\ln(\sqrt{\bar p C}N^{\rm coh}).
	\end{array}
	\end{equation}
	Letting $x=\ln(\sqrt{\bar p C}N^{\rm coh})$ and then $N^{\rm coh}=\frac{e^x}{\sqrt{\bar p C}}$, after some transformations, (\ref{eq26}) can be rewritten as
	\begin{equation}\label{eq27}
	\begin{array}{l}
	\frac{eTR_{\rm F}\sqrt{\bar p C}}{b}=(1+x)e^{x+1}.
	\end{array}
	\end{equation}
	By applying Lambert's $W$ function, $x$ can be expressed as
	\begin{equation}\label{eq28}
	\begin{array}{l}
	x=W_0\left(\frac{eTR_{\rm F}\sqrt{\bar p C}}{b}\right)-1.
	\end{array}
	\end{equation}
	After some transformations, we can obtain (\ref{eq24}). Meanwhile, we can obtain $f''(N^{\rm coh})<0$, which means that $f(N^{\rm coh})$ is concave. Thus $f(N^{\rm coh,*})$ is the maximum value.
\end{proof}
\textit{Remark 1:} We can observe from \textit{Proposition 1} that certain crucial parameters, such as the transmit power of the BS and channel conditions, have a notable influence on the number of coherent phase shifts. It is worth noting that $W_0(x)$ is a monotonically increasing function of $x$ when $x > 0$. As the transmit power $p_{\rm t}$ increases and the channel conditions improve, it becomes possible to reduce the delivery overhead by decreasing the number of elements adopting the coherent phase shift. Consequently, this reduction in delivery overhead leads to a shorter delivery time and an enhancement in the throughput.
\subsubsection{Feasible Conditions}
So far, we have obtained the closed-form solutions for the delivery time and the number of reflecting elements adopting the coherent phase shift. However, these solutions are not necessarily feasible. Thus, in the following, we will discuss the feasible conditions for $N^{\rm coh}$. Problem (\ref{eq20}) is feasible when the following conditions hold, i.e.,
\begin{equation}\label{eq29}
\begin{array}{l}
N^{{\rm coh},*}> \frac{(E^{\max}-Tp_{\rm t})R_{\rm F}}{(p-p_{\rm t})b},
\end{array}
\end{equation}
and
\begin{equation}\label{eq30}
\begin{array}{l}
\frac{2(TR_{\rm F}-bN^{{\rm coh},*})}{R_{\rm F}\ln(2)}\geq \frac{R^{\min}}{\ln(\sqrt{\bar p C}N^{{\rm coh},*})}.
\end{array}
\end{equation}
\begin{proof}
	(\ref{eq30}) is derived from constraint $C_2$, while (\ref{eq29}) is derived from $t$ in (\ref{eq22}) since $\frac{bN^{\rm coh}}{R_{\rm F}}>\frac{E^{\max}-Tp_{\rm t}}{p-p_{\rm t}}$ holds, which completes the proof.
\end{proof}

\subsubsection{Case 2 with $t=\frac{E^{\max}-Tp_{\rm t}}{p-p_{\rm t}}$} 
We can obtain the closed-form solution for $N^{\rm coh}$ by the following proposition. 

\textit{Proposition 2:} In the large SNR region, when $N^{\rm coh}>0$ holds, we have $N^{\rm coh,*}=\frac{(E^{\max}-Tp_{\rm t})R_{\rm F}}{(p-p_{\rm t})b}$. 
\begin{proof}
	In the large SNR region, the objective function can be approximated as $f(N^{\rm coh})\triangleq(T-\frac{E^{\max}-Tp_{\rm t}}{p-p_{\rm t}})\log_2(\bar pC(N^{\rm coh})^2)+o(1)$. Considering the leading-order term and taking the derivative of $f(N^{\rm coh})$ with respect to $N^{\rm coh}$ equal to 0, since $E^{\max}>Tp$, we obtain $f'(N^{\rm coh})=(T-\frac{E^{\max}-Tp_{\rm t}}{p-p_{\rm t}})\frac{2}{N^{\rm coh}\ln(2) }>0$. Thus, $f(N^{\rm coh})$ is an increasing function with respect to $N^{\rm coh}$. Then, the upper bound of $N^{\rm coh}$ is $\frac{TR_{\rm F}}{b}$  and $\frac{(E^{\max}-Tp_{\rm t})R_{\rm F}}{(p-p_{\rm t})b}$. Since $\frac{(E^{\max}-Tp_{\rm t})}{(p-p_{\rm t})}<T$, thus $N^{\rm coh,*}=\frac{(E^{\max}-Tp_{\rm t})R_{\rm F}}{(p-p_{\rm t})b}$, which completes the proof.
\end{proof}


\textit{Remark 2:} From \textit{Proposition 2}, the number of elements adopting the coherent phase shift depends mainly on the delivery bits for each reflecting element. A small quantization bit $b$ yields a larger number of elements adopting coherent phase shifts. This phenomenon indicates that when the quantization bit $b$ is small, the system will have a smaller delivery overhead. In such cases, it becomes feasible to employ a larger number of elements adopting coherent phase shifts to enhance the throughput.
\subsubsection{Feasible Conditions}
Similarly, we obtain the closed-form solution to the $N^{\rm coh,*}$, and then we discuss the feasible condition for $N^{\rm coh,*}$. Problem (\ref{eq20}) is feasible when the following condition holds, i.e.,
\begin{equation}\label{eq31}
\begin{array}{l}
N^{\rm coh,*}\geq \sqrt{\frac{2^{\frac{R^{\min}(p-p_{\rm t})}{Tp-E^{\max}}} }{\bar p C}}.
\end{array}
\end{equation}
It is worth noting that there is also a feasible condition $N^{\rm coh,*}\leq\frac{(E^{\max}-Tp_{\rm t})R_{\rm F}}{(p-p_{\rm t})b}$, which is derived from Eq. (\ref{eq22}), but the optimal solution must satisfy this condition, which thus can be omitted.

\vspace{-10pt}
\section{Simulation Results}\vspace{-5pt}
In this section, simulation results are provided to verify the effectiveness of the proposed algorithm. The BS and the RIS are respectively located at (0, 0) and (50, 10), while users are uniformly and randomly distributed in a circle centered at (50, 0) with a radius of 5 m. The path loss exponents $\alpha_{\rm B-R}$ and $\alpha_{\rm R-U}$ are set to 2.2 for the BS-RIS link and the RIS-user link. The small-scale fading follows Rician distribution with a Rician factor being 3. Other parameters are listed as Table I. The results of all figures are obtained over 100 channel realizations.
\begin{spacing}{1.00}
	\begin{table}[!t]
		\newcommand{\tabincell}[2]{\begin{tabular}{@{}#1@{}}#2\end{tabular}}
		\centering
		\tiny
			\caption{System Parameters.}\vspace{-10pt}
			\begin{tabular}{|c|c|c|c|c|c|c|c|}
				
				\hline
				
				\tabincell{c}{\textbf{Notation}} & \tabincell{c}{\textbf{Value}} & \tabincell{c}{\textbf{Notation}} & \tabincell{c}{\textbf{Value}}& \tabincell{c}{\textbf{Notation}} & \tabincell{c}{\textbf{Value}}& \tabincell{c}{\textbf{Notation}} & \tabincell{c}{\textbf{Value}}\\
				\hline
				
				\tabincell{c}{$M$} & \tabincell{c}{8} &  \tabincell{c}{$K$} & \tabincell{c}{4} &\tabincell{c}{$b$} & \tabincell{c}{1 bit} &\tabincell{c}{$E^{\max}$} & \tabincell{c}{1 J}\\			
				\hline
				
				\tabincell{c}{$N^{\rm coh}$} & \tabincell{c}{8} & \tabincell{c}{$N^{\rm fix}$} & \tabincell{c}{4}  &\tabincell{c}{$p$} & \tabincell{c}{20 dBm}&\tabincell{c}{$R_k^{\min}$} & \tabincell{c}{$1$ bps/Hz}\\
				\hline
				
				\tabincell{c}{$\delta_k^2$} & \tabincell{c}{-80 dBm} & \tabincell{c}{$R_{\rm F}$} & \tabincell{c}{[5,15] bps/Hz} &\tabincell{c}{$T$} & \tabincell{c}{1 S} &\tabincell{c}{$N$} & \tabincell{c}{12}\\
				\hline
				
		\end{tabular} \vspace{-20pt}
	\end{table}	
\end{spacing}
	

\begin{figure*}
		\vspace{-30pt}
	\centering
	\subfigure[]
	{
		\includegraphics[width=1.5in]{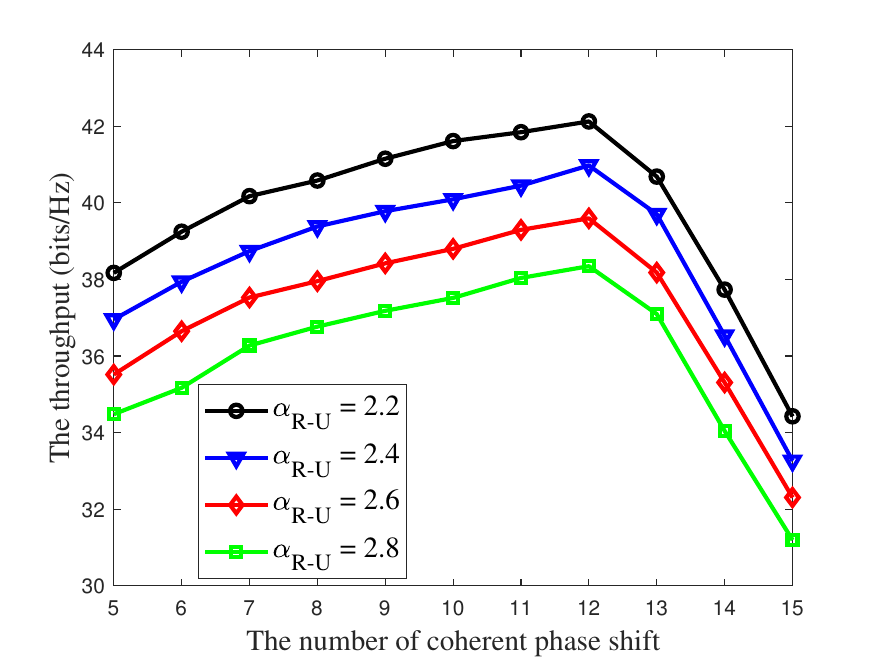} 
	}
	\subfigure[]
	{
		\includegraphics[width=1.5in]{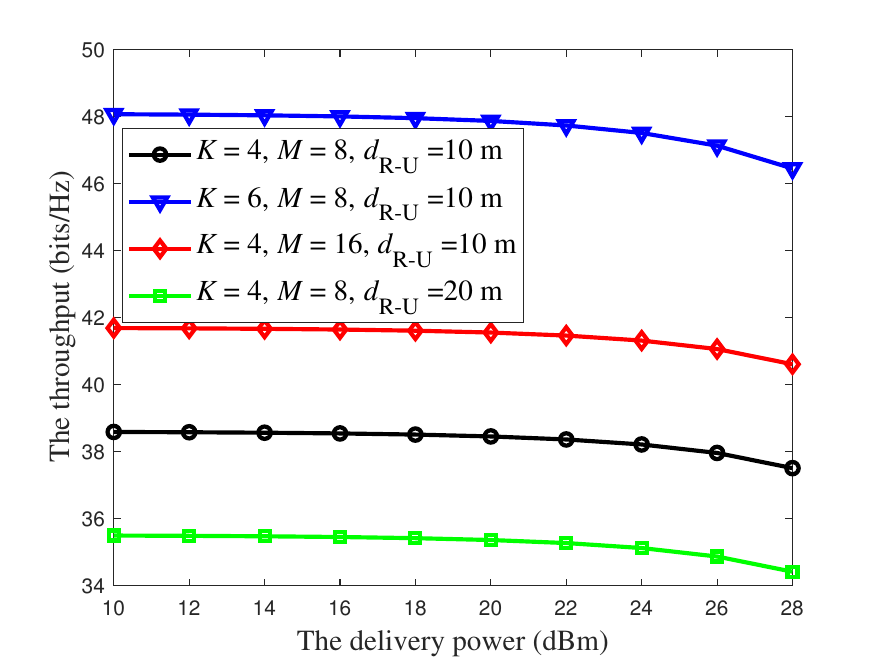} 
	}
	\subfigure[]
	{
		\includegraphics[width=1.5in]{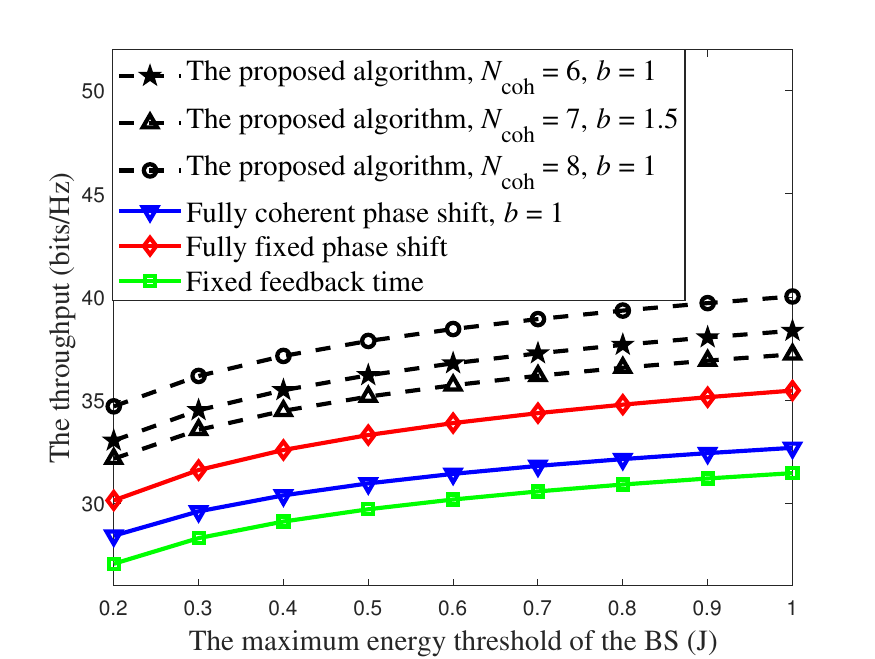} 
	}
	\DeclareGraphicsExtensions.\vspace{-10pt}
	\caption{(a) The throughput versus $N^{\rm coh}$. (b) The throughput versus $p$. (c) The throughput versus $E^{\max}$.}
	\label{fig2}\vspace{-20pt}
\end{figure*}

Fig. \ref{fig2}(a) illustrates the throughput versus the number of reflecting elements adopting coherent phase shift. We can observe that the throughput increases initially and then decreases as the number increases. This behavior can be attributed to the fact that, with a small number of elements, the delivery overhead during the phase-shift delivery phase is low, resulting in a shorter delivery time. However, as the number of elements grows larger, the delivery overhead increases, leading to an extended delivery time, which in turn reduces the transmission time and ultimately decreases the throughput. Notably, under the given parameter configuration, we also discover that the optimal number  is 12. Furthermore, it is worth noting that the throughput decreases as the path-loss exponents increase. This decrease is due to the greater signal attenuation during propagation, which results in a reduced throughput.

Fig. \ref{fig2}(b) shows the throughput versus the delivery power. We can observe that, with the increase in delivery power and distance between the RIS and the user, the throughput gradually decreases. This is because, according to $C_1$, the increase in the delivery power forces a reduction in the transmit power during the information transmission, and increasing the distance will result in the transmission signal experiencing more significant signal attenuation, and these factors will reduce the throughput. Besides, an increase in the number of users and the number of antennas can enhance the throughput. Undoubtedly, involving more users in the communication process results in a higher throughput. Additionally, the increase in the number of antennas provides a higher beamforming gain, all of which contribute to the improvement of the throughput.

Fig. \ref{fig2}(c) gives the throughput versus the maximum energy threshold of the BS under different algorithms. Without a doubt, the throughput increases with the increasing energy threshold since the increase in the energy threshold expands the feasible region of transmit power of the BS. As expected, the proposed algorithm outperforms other algorithms. When the number of quantization bit for each reflecting element increases, it leads to an increase in the delivery time, thereby reducing throughput. It is interesting to note that the performance of the fully coherent phase shift is worse than that of fully fixed phase shifts, which is different from previous works. This is because, although the fully coherent phase shift can maximize the SINR, it also significantly increases the delivery time, thus reducing the transmission time, and furthermore the throughput performance.

\vspace{-10pt}
\section{Conclusion}
In this letter, we investigate the  trade-off between the delivery overhead and the throughput. We focus on the hybrid phase shift mechanism in RIS-aided networks, which combines coherent and fixed phase shifts. In particular, the throughput is maximized by jointly optimizing the beamforming vector, the coherent phase shift, and the delivery time, which can be solved by an alternating optimization-based iterative algorithm by combining quadratic transformation and SCA. To gain more insights, we also derive the closed-form solution of the number of elements adopting the coherent phase shift in the large SNR region to guide the practical deployment of the RIS. Simulation results verify that the proposed algorithm achieves a good throughput-overhead trade-off.

\end{document}